\documentclass[11pt]{article}
\usepackage[utf8]{inputenc}
\usepackage[utf8]{inputenc}
\usepackage[T1]{fontenc}
\usepackage{amsfonts} 
\usepackage{url}
\date{\vspace{-5ex}}
\usepackage{amsmath}
\usepackage[a4paper, total={6in, 8in}]{geometry}
\usepackage{mathtools}
\numberwithin{equation}{section}
\usepackage{graphicx}           
\usepackage{adjustbox}
\usepackage{pdfpages}
\usepackage{float}
\usepackage{amssymb} 
\usepackage{amsmath}
\usepackage{amsfonts}
\usepackage{bm}
\usepackage{multirow}
\usepackage{amssymb}
\usepackage{booktabs}
\usepackage{natbib}
\usepackage{lipsum}
\usepackage{algorithm}
\usepackage{algorithmicx}
\usepackage{algpseudocode}
\usepackage{pdfpages}
\title{Time-to-event prediction for grouped variables using Exclusive Lasso}
\author{Dayasri Ravi, Andreas Groll}
\date{Department of Statistics, TU Dortmund University}

\begin{document}
\maketitle
\begin{abstract}
    The integration of high-dimensional genomic data and clinical data into time-to-event prediction models
has gained significant attention due to the growing availability of these
datasets. Traditionally, a Cox regression model is employed, concatenating various
 covariate types linearly. Given that much of the data may be
redundant or irrelevant, feature selection through penalization is often desirable.
A notable characteristic of these datasets is their organization into blocks
of distinct data types, such as methylation and clinical predictors, which requires
selecting a subset of covariates from each group due to high intra-group correlations.
For this reason, we propose utilizing Exclusive Lasso regularization in place of 
standard Lasso penalization. We apply our methodology to a real-life cancer dataset, demonstrating enhanced
survival prediction performance compared to the conventional Cox regression
model.
\end{abstract}
\section{Introduction}
In recent years, advancements in high-throughput genomic technologies have led to the availability of high-dimensional datasets, including DNA methylation, mRNA expression, and copy number variation, in addition to traditional clinical variables. These datasets may provide valuable information on the mechanisms of a certain disease, prompting the development of various methods to identify influential genomic and clinical characteristics for improved prognostic modeling.

A common objective in clinical research is the prediction of patient survival outcomes. The Cox proportional hazards (PH) model \citep{dr1972regression} is widely used for this purpose, as it not only facilitates survival prediction but also enables the assessment of the impact of predictor variables on survival. However, given the high-dimensional nature of genomic datasets, variable selection becomes a critical step in model construction. To address this, an $L_{1}$-penalized Cox model, such as the Lasso, is often employed to identify the most relevant features in time-to-event modeling \citep{tibshirani1997lasso}. 

Despite its effectiveness, this approach presents several limitations. First, standard Lasso-based methods do not inherently account for grouped variables, which is particularly relevant in genomic studies where genes are often organized in biological pathways. Ignoring such group structures may lead to suboptimal feature selection and loss of biologically meaningful information. Additionally, large sets of genomic features often overshadow low-dimensional clinical variables, such as tumor size and nodal status. This is a significant drawback, as clinicopathologic variables have been demonstrated to play a crucial role in oncological studies, and predictive performance improves when both clinical and genomic data are integrated \citep{ma2007supervised,herrmann2021large}. Finally, Lasso-based selection methods have been shown to produce a relatively high rate of false positives in certain settings, which may limit their reliability in time-to-event analysis, depending on the context. \citep{meinshausen2006high,zhao2006model}.

Several statistical methods have been proposed to incorporate grouped variables in the Cox PH model. Although Elastic Net Cox \citep{simon2011regularization} does not explicitly enforce group selection, it tends to select correlated variables together, unlike pure Lasso, which typically selects only one from a correlated set. This behavior results from its combination of $L_{1}$ and $L_{2}$-norm penalties. However, it does not guarantee that entire groups of variables will be retained or removed together. One of the most common approaches is Group Lasso \citep{kim2012analysis}, which applies $L_{2,1}$-norm regularization. This method enforces sparsity across groups using the $L_{1}$-norm while applying the $L_{2}$-norm within each group to regularize coefficients together. However, Group Lasso performs poorly when dealing with highly correlated groups, such as those found in multi-omics datasets. In such cases, it tends to select variables from only a few dominant groups, often overlooking smaller or lower-dimensional groups like clinical variables, which are essential for time-to-event prediction. This requires a method that can ensure the selection of variables for each group rather than selecting entire groups of variables. To overcome this challenge, Sparse Group Lasso was introduced \citep{simon2013sparse}, incorporating sparsity at both the group and individual variable levels. Another advancement in this area is the Integrative $L_{1}$-Penalized Regression with Penalty Factors (IPF-Lasso; \citealp{boulesteix2017ipf}), which allows for different penalty terms across variable groups, either based on prior knowledge or data-driven selection. However, none of these methods guarantee the selection of at least one variable from every group, which can lead to the exclusion of smaller yet important groups from the model. In biomedical studies, representing all relevant groups is important for gaining a complete understanding of underlying relationships.

To overcome this limitation, we propose the use of Exclusive Lasso regularization \citep{campbell2017within}. Exclusive Lasso encourages intra-group sparsity through the $L_{1}$-norm while promoting inter-group selection via the $L_{2}$-norm, ensuring that at least one variable is selected from each group. The properties of this regularization have been well studied in the literature \citep{campbell2017within,gregoratti2021exclusive}.

In our previous work, we demonstrated the superior performance of Exclusive Lasso over traditional Lasso in GLM settings with high within-group correlation (\citealp{ravi25}; note that a preliminary compact version of this work can also be found in \citealp{ravi2024optimizing}). However, this approach has not yet been transferred to time-to-event prediction. In this study, we extend Exclusive Lasso to the Cox PH model, introducing it as a practical alternative for selecting informative predictors from different groups. Our goal is to integrate these selected features into a sparse prediction model while ensuring that no group is overlooked.

We assess the performance of our proposed method by comparing it to traditional approaches that account for grouping effects, such as Elastic Net Cox, Sparse Group Lasso, and IPF-Lasso. Through extensive simulation studies, we show that our method consistently outperforms these alternatives across a range of scenarios. Additionally, we evaluate the practical applicability of our model by using it for survival prediction in a bladder cancer study. In addition to the standard prediction errors, we compare the biomarkers selected by each model and highlight the importance of Exclusive Lasso in selecting clinical and low-dimensional variables that other models fail to capture.

The remainder of the manuscript is structured as follows. Section~\ref{Ravi:Sec2}
introduces the Exclusive Lasso problem in the Cox PH framework. In Section~\ref{Ravi:Sec3}, we present the simulation scenarios and compare our method with other Lasso procedures. The applicability of our model is demonstrated in Section~\ref{Ravi:Sec4} using the aforementioned application example. Finally, Section~\ref{Ravi:Sec5} concludes.

\section{Methods} \label{Ravi:Sec2}
In this section, we first briefly review traditional methods for handling grouped predictors within the Cox PH framework and then introduce the Exclusive Lasso regularization in the Cox PH model.

Let $i = 1, \dots,n$ denote the observations (patients) in the cohort. For each patient, we observe $(t_i, \delta_i, \mathbf{x}_i)$, where $t_i$ is the event or censoring time for patient $i$,
$\delta_i$ is the censoring indicator, which is 1 if an event is observed and 0 if the observation is censored and $\mathbf{x}_i = (x_{i1}, \dots, x_{ip})^T$ is the vector of covariates associated with patient $i$.

The partial log-likelihood function for the Cox PH model \citep{dr1972regression}, with parameter vector $\boldsymbol{\beta}$, is given by:

\begin{equation}
   \ell(\boldsymbol{\beta}) = \sum_{i=1}^{n} \delta_i \left[ \mathbf{x}_i^T \boldsymbol{\beta} - \log \left( \sum_{l \in R(t_i)} \exp(\mathbf{x}_l^T \boldsymbol{\beta}) \right) \right] \; ,
   \label{Ravi:Equ1}
\end{equation}
where $R(t_i)$ represents the risk set at time $t_i$, which includes all individuals who are still at risk (i.e., uncensored and have not yet experienced the event) at the time of observation.

In high-dimensional scenarios, where the number of covariates $p$ by far exceeds the number of patients $n$, the estimation of the coefficients is typically performed by introducing a penalty term, $P(\boldsymbol{\beta})$, to the partial log-likelihood function. The estimation is then carried out by maximizing the penalized partial log-likelihood function, given by

\begin{equation}
    \ell_{\text{pen}}(\boldsymbol{\beta}) = \ell(\boldsymbol{\beta}) - \lambda P(\boldsymbol{\beta})\;, 
    \label{Ravi:Equ2}
\end{equation}
where $\lambda \geq 0$ is the penalty parameter.
The most common penalty term is the $L_1$-norm penalty, i.e.,\ $P(\boldsymbol{\beta}) = \sum_{k=1}^{p} \lvert \beta_k \rvert$. This penalty shrinks the coefficients towards zero and forces some of the coefficients corresponding to less important variables to be exactly zero, effectively performing variable selection.

We focus on variable selection in scenarios where the predictors are divided into predefined, disjoint groups. For instance, in the context of multi-omics data, the variables may include different types, such as genomics, epigenomics, and transcriptomics, in addition to clinical and pathological data. We assume that the indices of the true parameter vector $\boldsymbol{\beta}$ are divided into non-overlapping groups. Let
$\mathcal{G}$ be a collection of non-overlapping predefined groups {$\{1,\ldots,G\}$} of indices $g$ such that the union of all groups covers the entire set of indices, i.e.,
$$
\bigcup\limits_{g \in \mathcal{G}} = \{1,\ldots,p\}\,.
$$

\subsection*{Elastic Net} 
Elastic Net \citep{zou2005regularization} is a regularization method that combines the $L_1$ (Lasso) and $L_2$ (Ridge) penalties. The $L_1$ penalty encourages sparsity by shrinking some coefficients to zero, while the $L_2$ penalty promotes the inclusion of correlated variables in groups. This combination allows groups of correlated variables to be selected together. Elastic Net is particularly effective in situations where predictors are highly correlated within groups. The Elastic Net penalty is defined as :

\[
P(\boldsymbol{\beta}) = \alpha \sum_{k=1}^{p} \lvert \beta_k \rvert + \frac{1}{2}(1 - \alpha) \sum_{k=1}^{p} \beta_k^2 \;,
\]
where $\alpha \in (0,1)$ is the mixing parameter that controls the balance between the $L_1$ and $L_2$ penalties.

\subsection*{Sparse Group Lasso}
Sparse Group Lasso \citep{simon2013sparse} is another method that uses a combination of $L_{1}$- and $L_{2}$-norm penalties to encourage sparsity both across groups and within each group. The penalty term for Sparse Group Lasso is given by:

\[
P(\boldsymbol{\beta}) = (1 - \alpha) \sum_{g \in \mathcal{G}} \sqrt{p_g} \lVert \boldsymbol{\beta}_g \rVert_2 + \alpha \sum_{k=1}^{p} \lvert \beta_k \rvert \;,
\]
 where $\boldsymbol{\beta}_g$ represents the coefficients in group $g \in \mathcal{G}$, and $p_g$ is the number of variables in group $g$. This promotes group-wise selection, where all variables in a group are either included or excluded together. The second term is the Lasso penalty applied to individual coefficients, promoting sparsity at the level of individual predictors. When the parameter $\alpha = 0$, the Sparse Group Lasso reduces to the standard Group Lasso, and when $\alpha = 1$, it becomes the Lasso.
    
\subsection*{Integrative Lasso with penalty factors (IPF-Lasso)}
The Integrative Lasso with penalty factors (IPF-Lasso; \citealp{boulesteix2017ipf}) was introduced for prediction based on multi-omics datasets where there are several modalities (groups) of variables. The main idea of IPF-Lasso is to apply Lasso to each group and introduce penalty factors for different groups of variables, which can be selected according to the desired weighting of the groups or by cross-validation (CV). The IPF-Lasso penalty is defined as
$$
\sum_{g \in \mathcal{G}} \lambda_{g} \lVert \boldsymbol{\beta}_g \rVert_1 \;,
$$
where $\lambda_{g}$ is the penalty factor applied to the variables in group $g$. These penalty factors are chosen by CV via a grid search over a list of prespecified candidate vectors. However, this can be a time-consuming process. To avoid manually defining candidate vectors, we follow the procedure outlined in the Two-step IPF-Lasso \citep{schulze2017clinical}. In Step 1 of the process, before applying IPF-Lasso, a standard Lasso or Ridge regression is performed, and the arithmetic mean of the estimated coefficients can be considered as potential penalty parameters.

\subsection*{Exclusive Lasso}
The Exclusive Lasso \citep{campbell2017within} enforces structured sparsity by ensuring that at least one variable is selected from each predefined group. It combines $L_1$- and $L_2$-norm penalties, where the $L_1$ penalty within each group promotes the selection of informative variables, while the $L_2$-norm across groups prevents entire groups of coefficients from being set to zero. This approach ensures that even low-dimensional groups are represented while selecting only the most relevant variables from high-dimensional groups. The Exclusive Lasso penalty, which can be added to the Cox PH partial log-likelihood, is defined as:
\begin{equation}
    P(\boldsymbol{\beta}) = \frac{1}{2} \sum\limits_{{g\in\mathcal{G}}} \biggl(\sum\limits_{{k\in g}} \lvert \beta_{k} \rvert\biggr)^{2}.
    \label{Ravi:Equ3}
\end{equation}

The composite nature of the penalty term makes the estimation of the Exclusive Lasso problem challenging. Several strategies have been developed to tackle this challenge. One approach utilizes proximal point algorithms based on dual Newton methods \citep{lin2020adaptive}, while others employ iterative re-weighted techniques to refine the estimation process \citep{kong2014exclusive, sun2020correlated}. An alternative strategy reformulates the problem in a Lasso framework and applies a bisection algorithm, taking advantage of Lasso's piecewise linear properties \citep{sun2020correlated}. 

More recently, a fast optimization method leveraging the iterative shrinkage-thresholding algorithm (FISTA) has been proposed to improve computational efficiency \citep{huang2018exclusive}.
Another approach transforms the penalty into a differentiable one by applying a simple quadratic approximation, allowing it to be efficiently solved using a Newton-based algorithm \citep{ravi25}. However, to the best of our knowledge, none of these methods have been extended to time-to-event prediction yet. 

To address this challenge, we employ the coordinate descent method with soft-thresholding. As highlighted by \citet{campbell2017within}, the Exclusive Lasso penalty cannot be expressed as a sum of separable functions, i.e.,

\begin{equation*}
P(\boldsymbol{\beta}) \neq \sum_{j=1}^{p} P_{j}(\beta_{j}).
\end{equation*}

This implies that a simultaneous update of all variables is not feasible. Instead, estimation is performed using a coordinate descent algorithm, where each variable is updated sequentially while keeping the others fixed.  

Our approach adopts a coordinate descent framework tailored for Group Lasso regularization \citep{yuan2006model}, which has been shown to be efficient in solving high-dimensional problems. The proposed algorithm is presented in Algorithm~\ref{Ravi:Alg1}.  

The gradient component for covariate \( j \) in the Cox PH model from Equation~\eqref{Ravi:Equ1} is defined as:
\[
\hat{r}_j = \sum_{i=1}^{n} \delta_i \left[ x_{ij} - \frac{\sum\limits_{l \in R(t_i)} x_{lj} \exp(\mathbf{x}_l^T \boldsymbol{\beta})}{\sum\limits_{l \in R(t_i)} \exp(\mathbf{x}_l^T \boldsymbol{\beta})} \right],
\]
where \( x_{lj} \) represents the observed value of covariate \( j \) for individual \( l \), and \( R(t_i) \) denotes the risk set at time \( t_i \). This formulation ensures that only individuals for whom \( \delta_i = 1 \) (i.e., those who experience an event) contribute to the estimation of \( \beta_j \).

We apply coordinate descent to maximize the penalized partial log-likelihood defined in Equation~\eqref{Ravi:Equ2}, using the soft-thresholding operator \( S(z, \lambda) \) given by:

\begin{equation}
S(z, \lambda) = \operatorname{sign}(z) \max(|z| - \lambda, 0).
\label{Ravi:Equ4}
\end{equation}

The soft-thresholding operator shrinks coefficients toward zero by subtracting a threshold~$ \lambda$, setting them exactly to zero when their magnitude falls below this threshold. This encourages sparsity and facilitates automatic variable selection.

Since we update one coefficient \( \beta_j \) at a time while holding others fixed, we use \( g \setminus j \) to denote the set of indices in group \( g \) excluding index \( j \). The corresponding penalty term for \( \beta_j \) is then updated as:

\[
\tilde{P}_j = \lambda \sum_{l \in g \setminus j} |\beta_l|.
\]

This penalty encourages competition among variables within the same group, allowing only a few features to get selected. It promotes sparsity by shrinking coefficients, especially when the penalty is large. As a result, \( \beta_j \) is pushed toward zero when other covariates in the group have large values, reducing redundancy among correlated variables.

Furthermore, we refer readers to Theorem 4 of \citet{campbell2017within}, which provides proof that the Exclusive Lasso coordinate descent algorithm converges to the global minimum.

\begin{figure}[t]\centering
\includegraphics[width=1\textwidth]{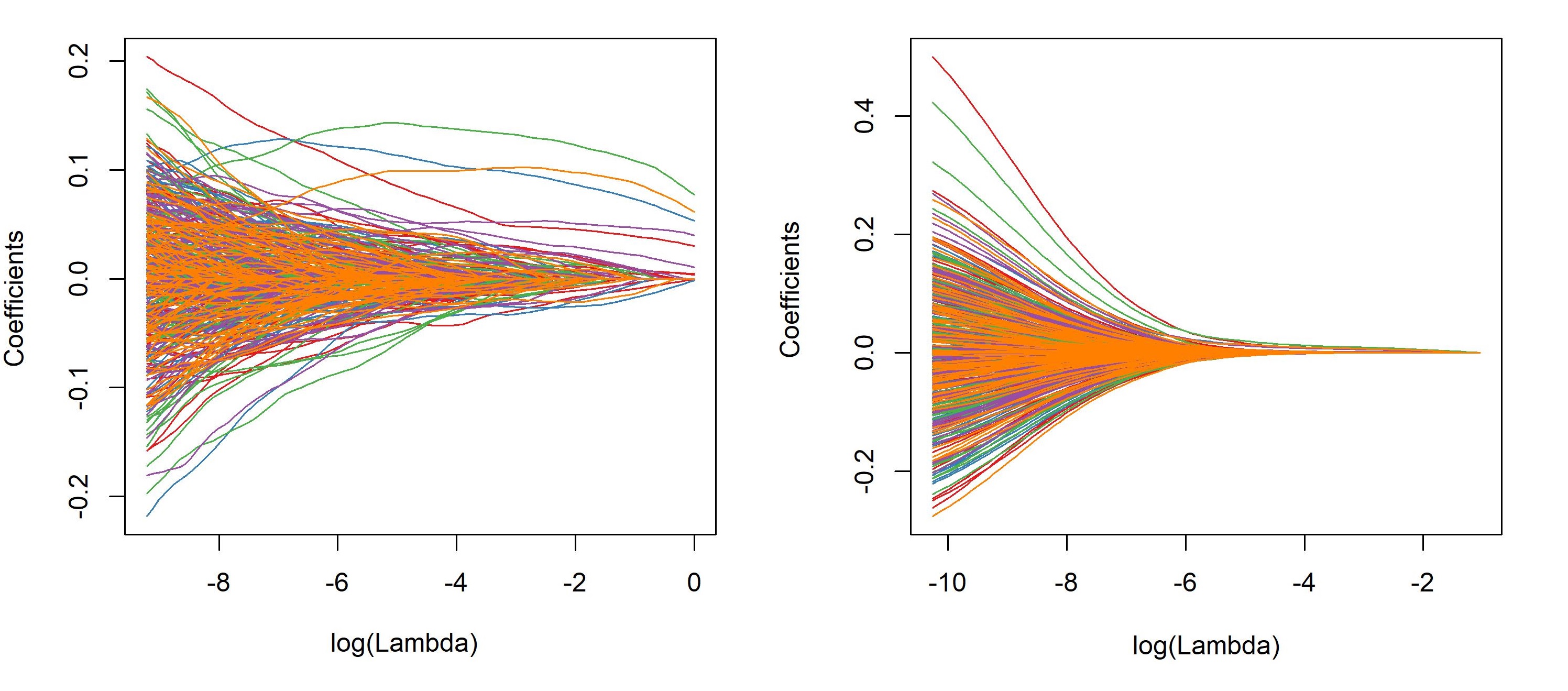}
\caption{Regularization paths for Exclusive Lasso (\emph{left}) and Lasso (\emph{right}) from a simulation study, where variables are evenly distributed into five groups (shown in distinct colors), with each group containing one true signal variable. In the Exclusive Lasso model, the true variables remain active unless all other variables in their group shrink to zero. In contrast, the Lasso model selects variables without considering the group structure, allowing multiple variables from the same group to be included.
}
\label{Ravi:Fig0}
\end{figure}

\begin{algorithm}[!h]
    \caption{Exclusive Lasso coordinate descent for Cox PH model}
    \label{Ravi:Alg1}\begin{algorithmic}[1]
        \Require Initial coefficients \( \boldsymbol{\beta}^0 \in \mathbb{R}^p \), tolerance \( \epsilon > 0 \), regularization parameter \( \lambda > 0 \), group structure \( \mathcal{G} \), design matrix \( \mathbf{X} \in \mathbb{R}^{n \times p} \), event times \( \mathbf{T} \in \mathbb{R}^n \), event indicators \( \boldsymbol{\delta} \in \{0,1\}^n \)
\State \textbf{Precompute:} Sort the data in increasing order of event times \( \mathbf{T} \)
        \While{$||\boldsymbol{\beta}^{(k+1)} - \boldsymbol{\beta}^{(k)}|| > \epsilon$}
            \ForAll{groups \( g \in \mathcal{G} \)}
                \ForAll{features \( j \in g \)}
                    \State Compute gradient component for covariate $j$  :
                    \[
                    \hat{r}_j = \sum_{i=1}^{n} \delta_i \left[ x_{ij} - \frac{\sum\limits_{l \in R(t_i)} x_{lj} \exp(\mathbf{x}_l^T \boldsymbol{\beta})}{\sum\limits_{l \in R(t_i)} \exp(\mathbf{x}_l^T \boldsymbol{\beta})} \right],
                    \]
                    \State Compute Exclusive Lasso Penalty terms:
                    \[
                    \tilde{P}_j = \lambda \sum_{l \in g \setminus \{j\}} |\beta_l|
                    \]
                    \State Compute Hessian approximation:
                    \[
                    H_j = \sum_{i=1}^{n} \delta_i x_{ij}^2
                    \]
                    \State Update \( \beta_j \) using soft-thresholding from Equation~\eqref{Ravi:Equ4}:
                    \[
                    \beta_j^{(k+1)} = S \left( \frac{\hat{r}_j}{H_j+\lambda}, \frac{\tilde{P}_j}{H_j+\lambda} \right)
                    \]
                    
                \EndFor
            \EndFor
        \EndWhile
        \State \Return \( \hat{\boldsymbol{\beta}} \)
    \end{algorithmic}
\end{algorithm}
\noindent Figure~\ref{Ravi:Fig0} displays the regularization paths for both Exclusive Lasso (left) and Lasso (right). The variables are divided into five distinct groups, with each group containing exactly one signal variable and the rest being noise. The signal variables are highlighted using different colors to distinguish their respective groups. Exclusive Lasso encourages within-group sparsity, driving most coefficients to zero while retaining only one active variable per group. As a result, it maintains exactly five active variables, one from each group, even at large values of \( \lambda \). In contrast, Lasso applies shrinkage without regard to group structure and may eliminate informative variables or retain multiple variables from the same group.

\section{Simulations} \label{Ravi:Sec3}
In this section, we present a detailed simulation study to evaluate the performance of our method across different scenarios.

\subsection{Setting} \label{Ravi:Sec3.1}
We simulate \( n = 500 \) observations and \( p = 500 \) variables from a multivariate Gaussian distribution with a Toeplitz covariance matrix \( \pmb{\Sigma} \), where the entries \( \Sigma_{i,j} = 0.6^{\lvert{i-j}\rvert} \) for variables in the same group, and \( \Sigma_{i,j} = 0.3^{\lvert{i-j}\rvert} \) for variables in different groups. Altogether, we use a moderate correlation, resulting in a high correlation within groups and a low correlation between groups. Event times are simulated using a Cox PH model framework, where the hazard function depends on a baseline hazard and a linear combination of the predictors.
 We consider a baseline median event time of eight years. The true coefficients are drawn from a uniform distribution between 0.5 and 1.5. Independent censoring times are simulated using an exponential distribution with a rate of 0.02, assuming constant censoring hazard over time. 

We assume that the variables are divided into five groups and consider three different simulation scenarios for grouping them. The total number of signal variables in all three scenarios is set to 5, 10, and 20, respectively.  

Table~\ref{Ravi:Tab1} provides a detailed overview of the grouping structure used across all three scenarios.
 In the first scenario, we allocate an equal number of variables to each group, with each group containing 100 variables. This represents an ideal setting, as Exclusive Lasso is expected to perform well when at least one signal variable is present in each group.  

In the second and third scenarios, we introduce unequal group sizes. In Scenario 2, the variables are distributed as (15, 20, 85, 180, 200), while in Scenario 3, the distribution is (5, 295, 10, 90, 100). Additionally, the signal variables are also unequally distributed among the groups.   

\begin{table}[h]
    \centering
    
    \renewcommand{\arraystretch}{1.2} 
    \begin{adjustbox}{valign=c, max width=\textwidth}
    \begin{tabular}{|p{2.5cm}|p{1.5cm}|p{1.5cm}|p{1.5cm}|p{2.5cm}|p{1.5cm}|p{1.5cm}|p{1.5cm}|p{2.5cm}|p{1.5cm}|p{1.5cm}|p{1.5cm}|}
    \hline
    \multicolumn{4}{|c|}{Scenario 1} & \multicolumn{4}{c|}{Scenario 2} & \multicolumn{4}{c|}{Scenario 3} \\ \hline
    Variables per group & \multicolumn{3}{c|}{True variables per group} & Variables per group & \multicolumn{3}{c|}{True variables per group} & Variables per group & \multicolumn{3}{c|}{True variables per group} \\ \hline
    100 & 1 & 2 & 4 & 15 & 1 & 1 & 2 & 5 & 1 & 1 & 2 \\ \hline
    100 & 1 & 2 & 4 & 20 & 1 & 2 & 2 & 295 & 1 & 2 & 6 \\ \hline
    100 & 1 & 2 & 4 & 85 & 1 & 1 & 1 & 10 & 1 & 1 & 4 \\ \hline
    100 & 1 & 2 & 4 & 180 & 1 & 4 & 10 & 90 & 1 & 2 & 6 \\ \hline
    100 & 1 & 2 & 4 & 200 & 1 & 2 & 5 & 100 & 1 & 4 & 2 \\ \hline
    \end{tabular}
    \end{adjustbox}
    \caption{Description of grouping structure of both signal and noise variables across three simulation scenarios. }
    \label{Ravi:Tab1}
\end{table}
We simulate an independent validation dataset consisting of \( n = 500 \) observations to evaluate model performance. The penalty parameter \( \lambda \) is selected via 5-fold CV for all models. For the IPF-Lasso, we choose the value of \( \lambda \) by maximizing the cross-validated predictive log-likelihood via a 5-fold CV with 10 repeats. Although we initially aimed to assess our simulations using Sparse Group Lasso with a cross-validated mixing parameter~\( \alpha \), this approach significantly increased memory and time requirements. Thus, we fix \( \alpha = 0 \) to maintain computational efficiency comparable to other models. This choice corresponds to the standard Group Lasso and avoids using the default \( \alpha = 0.95 \) recommended by the authors of the \texttt{SGL} \texttt{R} package \citep{SGL}.

We report the results using variable selection accuracy, defined as the proportion of true positives and true negatives among all variables, along with the F1 score, false discovery rate (FDR), and integrated Brier score (IBS).
The F1 score \citep{van1979information} is defined as the harmonic mean of precision and recall, accounting for both false positives and false negatives. The metric ranges from 0 to 1, with larger values indicating a better balance between precision and recall. The Brier score \citep{graf1999assessment} at a given time point $t$ represents the average squared distances between the observed event status and the predicted survival probability. The IBS provides an overall performance of the model by integrating the Brier score at all available time points. A lower value of IBS is desired as it indicates that the model's predicted probabilities are closer to the true probabilities across all available time points.
\begin{table}[t]
\centering
\begin{adjustbox}{valign=c, max width=\textwidth}
\begin{tabular}{llrrrr}
  \toprule
No. of true variables & Metric & Elastic Net & Exclusive Lasso & Group Lasso & IPF \\ 
  \midrule
\multirow{4}{*}{5} & Selection Accuracy & 0.86 (0.001) & \textbf{0.99 (0.000)} & 0.01 (0.000) & 0.93 (0.001) \\ 
  & F1 score & 0.13 (0.001) & \textbf{0.67 (0.007)} & 0.02 (0.000) & 0.23 (0.002) \\ 
  & False discovery rate & 0.93 (0.000) & \textbf{0.49 (0.010)} & 0.99 (0.000) & 0.87 (0.001) \\ 
  & Integrated Brier score & 0.56 (0.001) & \textbf{0.42 (0.000)} & 0.44 (0.001) & 0.53 (0.000) \\ 
\hline
\multirow{4}{*}{10} & Selection Accuracy & 0.83 (0.000) & \textbf{0.98 (0.000)} & 0.02 (0.000) & 0.90 (0.001) \\ 
  & F1 score & 0.19 (0.000) & \textbf{0.65 (0.002)} & 0.04 (0.000) & 0.25 (0.001) \\ 
  & False discovery rate & 0.89 (0.000) & \textbf{0.52 (0.003)} & 0.98 (0.000) & 0.85 (0.001) \\ 
  & Integrated Brier score & 0.61 (0.001) & \textbf{0.43 (0.000)} & 0.47 (0.000) & 0.52 (0.001) \\ 
\hline
\multirow{4}{*}{20} & Selection Accuracy & 0.81 (0.001) & \textbf{0.91 (0.000)} & 0.04 (0.000) & 0.89 (0.001) \\ 
  & F1 score & 0.30 (0.001) & \textbf{0.47 (0.001)} & 0.08 (0.000) & 0.42 (0.001) \\ 
  & False discovery rate & 0.83 (0.001) & \textbf{0.69 (0.001)} & 0.96 (0.000) & 0.74 (0.001) \\ 
  & Integrated Brier score & 0.65 (0.000) & \textbf{0.43 (0.000)} & 0.49 (0.000) & 0.55 (0.000) \\ 
   \bottomrule
\end{tabular}
\end{adjustbox}
\caption{Average performance metrics (standard errors in brackets) for different models across varying numbers of signal variables in Scenario 1 over 100 iterations; best-performing modeling approach per setting in bold font.} 
\label{Ravi:Tab2}
\end{table}

\begin{figure}[t]\centering
\includegraphics[width=1.05\textwidth]{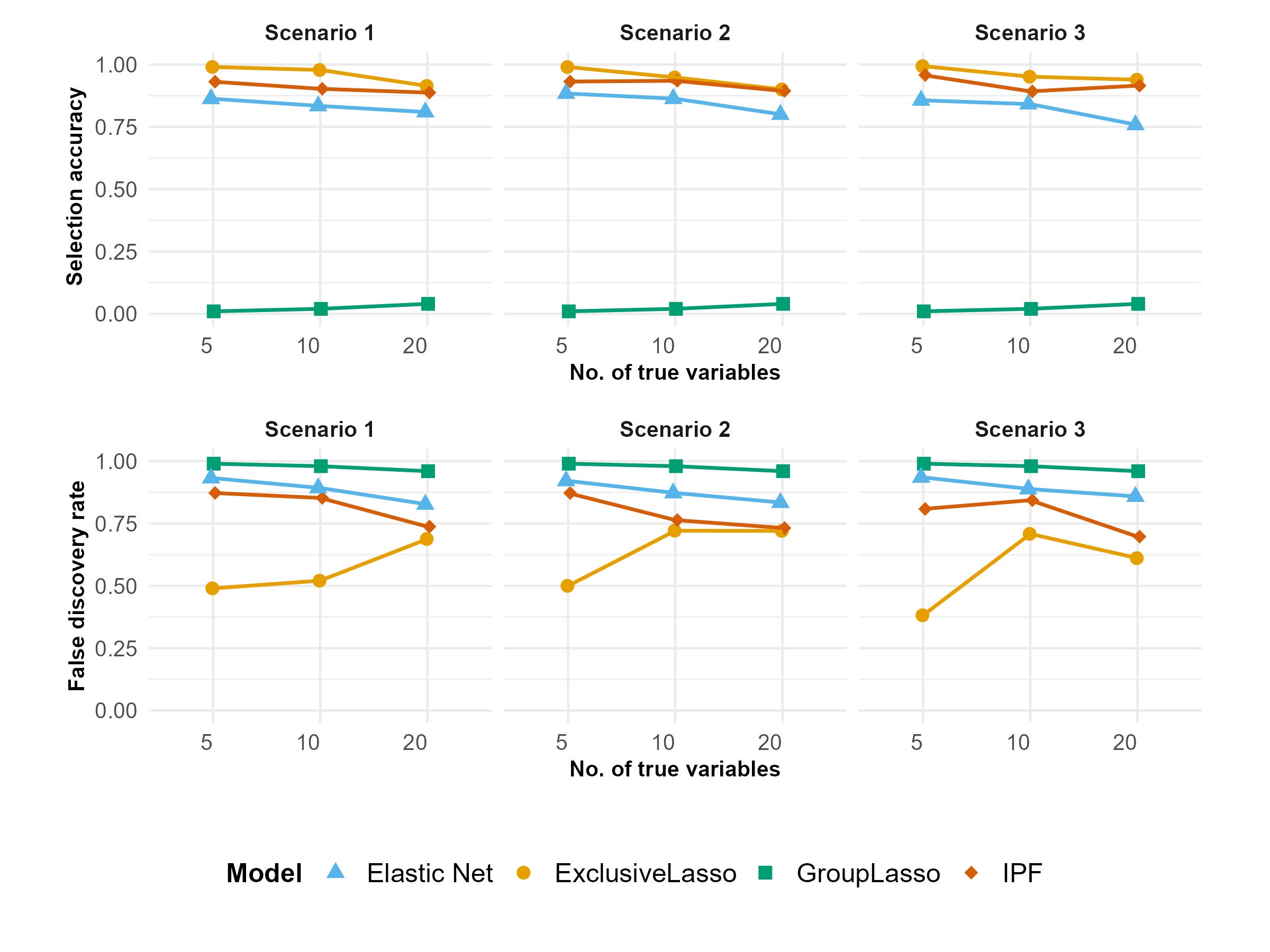}
\caption{Selection accuracy and false discovery rate across three scenarios for varying numbers of signal variables. The results are averaged over 100 randomly generated datasets.}
\label{Ravi:Fig1}
\end{figure}

\subsection{Results}
We compare our proposed extension of Exclusive Lasso for Cox PH models with the models described in Section~\ref{Ravi:Sec2}. We use the implementations available in the \textsf{R} packages: \texttt{glmnet} \citep{Coxnet,glmnet} for Elastic Net Cox, \texttt{SGL} \citep{SGL} for Group Lasso, and \texttt{ipflasso} \citep{ipflasso} for IPF-Lasso. We generate 100 random test datasets and report the average selection accuracy, F1 score, false discovery rate, and IBS for different numbers of true variables across Scenarios 1, 2, and 3 in Tables~\ref{Ravi:Tab2}--\ref{Ravi:Tab4}, respectively. Figure~\ref{Ravi:Fig1} visually represents the average selection accuracy and false discovery rate over 100 random replications. Exclusive Lasso demonstrates the largest selection accuracy and F1 score across all three simulation scenarios.

For most models, except Group Lasso, accuracy decreases as the number of signal variables increases. Group Lasso, on the other hand, shows a slight improvement with more signal variables, as it tends to select all variables within a group. However, despite this slight improvement, its overall performance remains quite poor.

Exclusive Lasso's performance declines more significantly in Scenarios 2 and 3, where variable allocation is more random (see Table~\ref{Ravi:Tab3}). As the number of signal variables increases, its accuracy becomes comparable to that of IPF-Lasso. In contrast, Elastic Net maintains consistent performance across all scenarios. This is because Elastic Net does not explicitly account for grouping, so variations in the number of variables per group do not impact its selection process.

Regarding the false discovery rate (FDR), Exclusive Lasso is the only model where FDR increases as the number of signal variables rises. For the other models, FDR decreases with more signal variables. This occurs because, as the number of variables or groups increases, Exclusive Lasso tends to select variables from every group, even if some groups contain only non-informative variables. This behavior is especially evident in Scenario 3, where signal variables are not evenly distributed among groups (see Table~\ref{Ravi:Tab4}). However, despite this increase in FDR, Exclusive Lasso still maintains a lower false discovery rate than the other models.

However, we do not observe significant differences in IBS scores across different scenarios or varying numbers of true variables. This could be due to the high correlation among variables, allowing the models to produce survival probability estimates that remain close to the true values despite differences in selection accuracy. As a result, IBS scores remain relatively stable across all models. Nonetheless, Exclusive Lasso exhibits slightly better predictive performance compared to the other models.

Overall, we find that Exclusive Lasso outperforms the other models in scenarios where variables are highly correlated and grouped. While its performance declines slightly in randomly allocated scenarios, it still remains superior to the other models. The only model that comes close to Exclusive Lasso is IPF-Lasso, and even then, only when the number of signal variables is large.

\begin{table}[t]
\centering
\begin{adjustbox}{max width=\textwidth}
\begin{tabular}{llrrrr}
  \toprule
No. of true variables & Metric & Elastic Net & Exclusive Lasso & Group Lasso & IPF \\ 
  \midrule
\multirow{4}{*}{5} & Selection Accuracy & 0.88 (0.000) & \textbf{0.99 (0.000)} & 0.01 (0.000) & 0.93 (0.000) \\ 
  & F1 score & 0.15 (0.000) & \textbf{0.67 (0.000)} & 0.02 (0.000) & 0.23 (0.001) \\ 
  & False discovery rate & 0.92 (0.000) & \textbf{0.50 (0.000)} & 0.99 (0.000) & 0.87 (0.000) \\ 
  & Integrated Brier score & 0.55 (0.001) & \textbf{0.43 (0.000)} & 0.49 (0.001) & 0.53 (0.001) \\ 
\hline
\multirow{4}{*}{10} & Selection Accuracy & 0.86 (0.001) & \textbf{0.95 (0.000)} & 0.02 (0.000) & 0.94 (0.001) \\ 
  & F1 score & 0.23 (0.001) & \textbf{0.44 (0.001)} & 0.04 (0.000) & 0.38 (0.003) \\ 
  & False discovery rate & 0.87 (0.001) & \textbf{0.72 (0.001)} & 0.98 (0.000) & 0.76 (0.002) \\ 
  & Integrated Brier score & 0.62 (0.001) & \textbf{0.43 (0.000)} & 0.51 (0.001) & 0.58 (0.001) \\ 
\hline
\multirow{4}{*}{20} & Selection Accuracy & 0.80 (0.001) & \textbf{0.90 (0.000)} & 0.04 (0.000) & 0.89 (0.000) \\ 
  & F1 score & 0.29 (0.001) & \textbf{0.43 (0.000)} & 0.08 (0.000) & 0.42 (0.000) \\ 
  & False discovery rate & 0.83 (0.000) & \textbf{0.72 (0.000)} & 0.96 (0.000) & 0.73 (0.000) \\ 
  & Integrated Brier score & 0.65 (0.001) & \textbf{0.43 (0.000)} & 0.51 (0.000) & 0.53 (0.000) \\ 
   \bottomrule
\end{tabular}
\end{adjustbox}
\caption{Average performance metrics (standard errors in brackets) for different models across varying numbers of signal variables in Scenario~2 over 100 iterations; best-performing modeling approach per setting in bold font.} 
\label{Ravi:Tab3}
\end{table}

\begin{table}[h!]
\centering
\begin{adjustbox}{max width=\textwidth}
\begin{tabular}{llrrrr}
  \toprule
No. of true variables & Metric & Elastic Net & Exclusive Lasso & Group Lasso & IPF \\ 
  \midrule
\multirow{4}{*}{5} & Selection Accuracy & 0.86 (0.000) & \textbf{0.99 (0.000)} & 0.01 (0.000) & 0.96 (0.000) \\ 
  & F1 score & 0.12 (0.000) & \textbf{0.76 (0.007)} & 0.02 (0.000) & 0.32 (0.001) \\ 
  & False discovery rate & 0.94 (0.000) & \textbf{0.38 (0.007)} & 0.99 (0.000) & 0.81 (0.001) \\ 
  & Integrated Brier score & 0.56 (0.001) & \textbf{0.44 (0.000)} & 0.46 (0.001) & 0.54 (0.001) \\ 
\hline
\multirow{4}{*}{10} & Selection Accuracy & 0.84 (0.001) & \textbf{0.95 (0.001)} & 0.02 (0.000) & 0.89 (0.001) \\ 
  & F1 score & 0.20 (0.001) & \textbf{0.45 (0.003)} & 0.04 (0.000) & 0.27 (0.001) \\ 
  & False discovery rate & 0.89 (0.000) & \textbf{0.71 (0.002)} & 0.98 (0.000) & 0.84 (0.001) \\ 
  & Integrated Brier score & 0.61 (0.001) & \textbf{0.43 (0.000)} & 0.47 (0.001) & 0.61 (0.001) \\ 
\hline
\multirow{4}{*}{20} & Selection Accuracy & 0.76 (0.001) & \textbf{0.94 (0.000)} & 0.04 (0.000) & 0.92 (0.000) \\ 
  & F1 score & 0.25 (0.001) & \textbf{0.54 (0.002)} & 0.08 (0.000) & 0.45 (0.001) \\ 
  & False discovery rate & 0.86 (0.000) & \textbf{0.61 (0.002)} & 0.96 (0.000) & 0.70 (0.001) \\ 
  & Integrated Brier score & 0.68 (0.001) & \textbf{0.43 (0.000)} & 0.49 (0.000) & 0.53 (0.000) \\ 
   \bottomrule
\end{tabular}
\end{adjustbox}
\caption{Average performance metrics (standard errors in brackets) for different models across varying numbers of signal variables in Scenario~3 over 100 iterations; best-performing modeling approach per setting in bold font.} 
\label{Ravi:Tab4}
\end{table}

\section{Application}\label{Ravi:Sec4}
Next, we apply our proposed method to a real-world gene expression dataset. The regularization parameter $\lambda$ is tuned via CV for all models.

Bladder cancer (BC) is one of the most commonly diagnosed urinary cancers worldwide, with its incidence steadily increasing each year. This rise may be linked to factors such as tobacco use and an aging population. Although the 5-year survival rate for BC is relatively high at 77\%, the recurrence rate remains a significant concern. Beyond genetic signatures, numerous risk factors contribute to BC development, including gender, smoking pattern, and occupational exposure to carcinogens \citep{cumberbatch2018epidemiology}. Therefore, it is crucial to incorporate both clinical risk factors and sensitive biomarkers when predicting overall survival in bladder cancer patients.

\begin{figure}[t]\centering
\includegraphics[width=\textwidth]{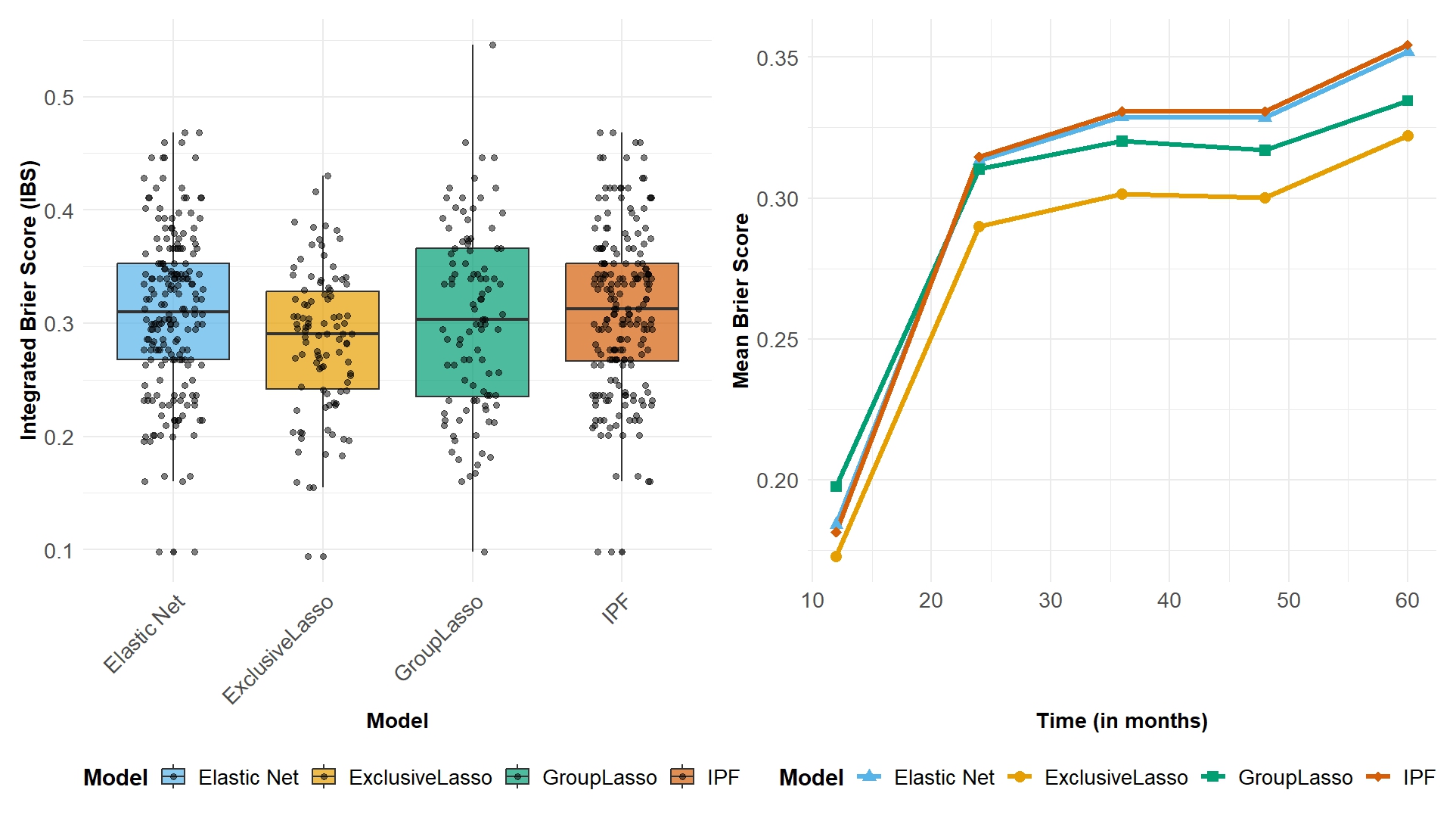}
\caption{Prediction errors for Bladder gene expression study. \emph{Left}: Boxplots of the integrated Brier score evaluated up to 60 months across 100 random training-test-data splits. \emph{Right}: Mean Brier scores calculated at 10, 20, 30, 40, 50, and 60 months, averaged over 100 random training-test-data splits.}
\label{Ravi:Fig2}
\end{figure}
\begin{figure}[h!]\centering
\includegraphics[width=0.97\textwidth]{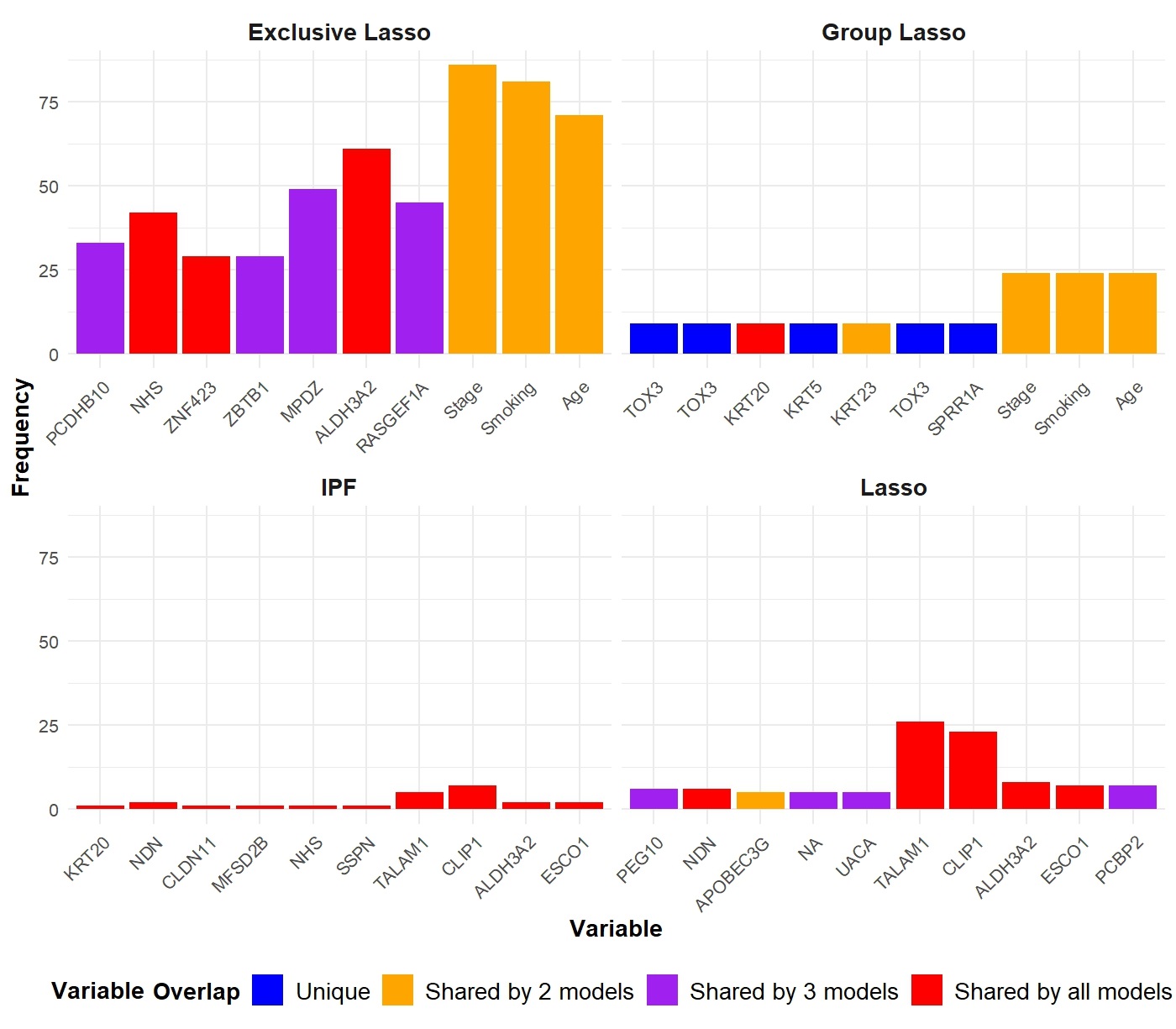}
\caption{The top 10 most frequently selected variables by the different models on the training set of the Bladder cancer gene expression study for 100 random subsamples.}
\label{Ravi:Fig3}
\end{figure}

\begin{figure}[ht]\centering
\includegraphics[width=0.9\textwidth]{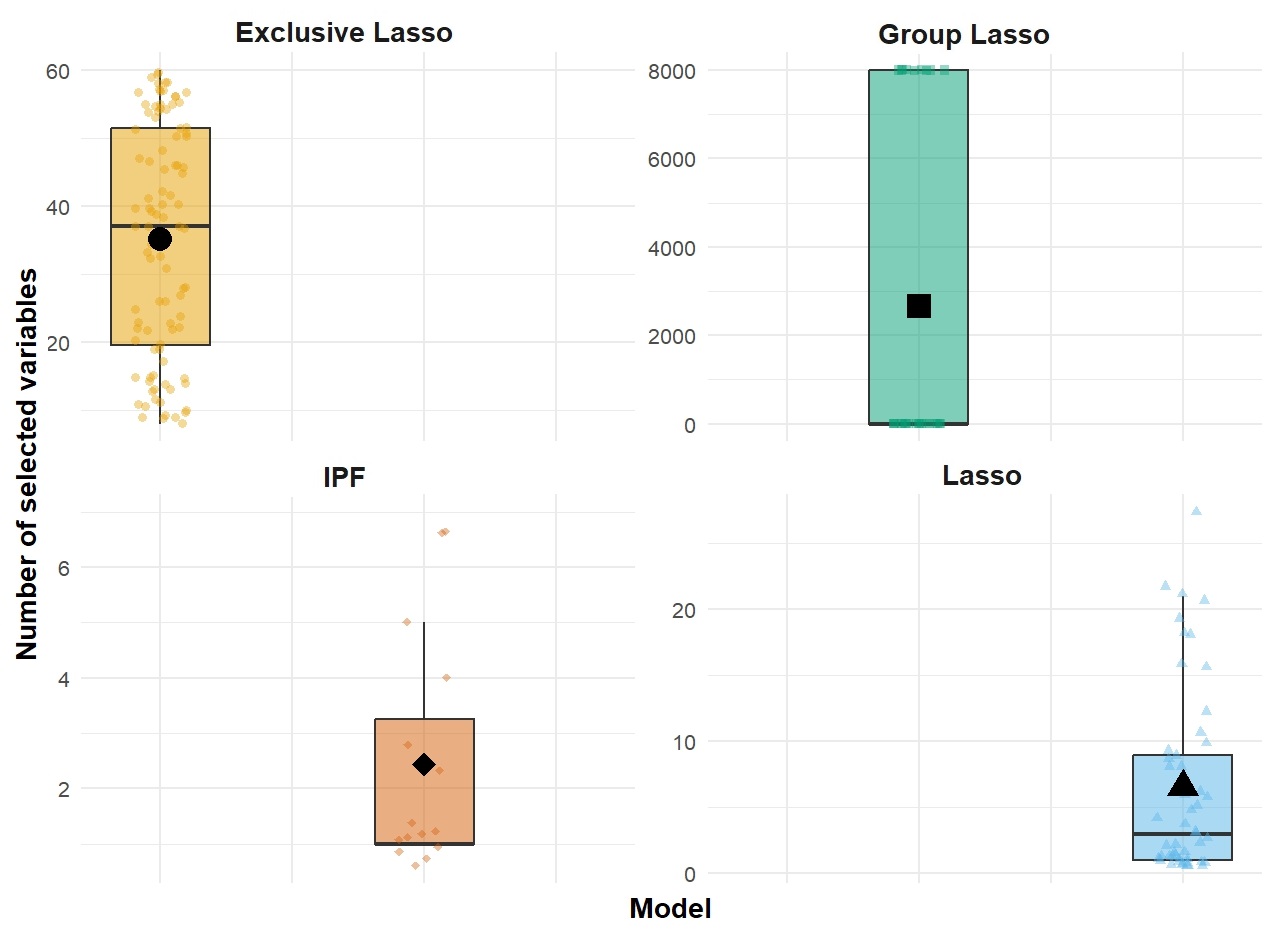}
\caption{Average number of selected variables across 100 random subsamples for each model in the Bladder cancer gene expression study. The black pointers are the average numbers of selected variables.}
\label{Ravi:Fig4}
\end{figure}

We analyze the BC dataset retrieved from the Gene Expression Omnibus (GEO) database (URL: \url{https://www.ncbi.nlm.nih.gov/geo/}) using the “GEOquery” \texttt{Bioconductor} \textsf{R} package, with the GEO accession GSE31684 \citep{riester2012combination}. The dataset includes gene expression data for 54,675 genes from 93 patients. For data preprocessing, we apply a variance filter to select genes with high variance, as previous studies have shown that using a variance filter before fitting a regularized Cox PH regression model can improve the performance of regularized Cox regression models and lead to stable feature selection \citep{bommert2022benchmark}. We categorize the variables into two groups: clinical and gene expression. The clinical group includes age (in years), tumor stage (Ta/T1, T2, T3, T4), nomogram score, and packs smoked per year. The data is split into training (70\%) and testing (30\%) sets, and this process is repeated for 100 times.  For each split, we compute prediction errors on the test set and identify the top 10 most frequently selected variables by each model on the training data. This frequency-based approach is motivated by the principle of stability selection \citep{meinshausen2010stability}, which aims to identify variables that are consistently selected across multiple resampled datasets, thereby improving the robustness and reliability of variable selection in high-dimensional settings.

We report the Brier scores computed up to 5 years for all the models discussed in Section~\ref{Ravi:Sec2}. From Figure~\ref{Ravi:Fig2}, we observe that Exclusive Lasso yields the lowest IBS curve. Although there is no substantial difference in the IBS across models, the right plot in Figure~\ref{Ravi:Fig2} shows that Exclusive Lasso consistently gives the lowest mean Brier score at each time point. Elastic Net and IPF-Lasso perform almost identically. Group Lasso shows larger Brier scores for events between 10 and 20 months, but beyond that, its Brier score is lower than that of Elastic Net and IPF-Lasso.

Figure~\ref{Ravi:Fig3} displays the top 10 most frequently selected variables by all models. We observe that the clinical variables ``Stage'', ``Smoking'', and ``Age'' are selected more frequently by Exclusive Lasso than by any other model. In 100 iterations, the variable ``Stage'' is selected more than 80 times. Despite being part of a low-dimensional clinical group, Exclusive Lasso consistently selects at least one variable from this group, while other models tend to ignore these variables completely. These variables are only selected by Group Lasso in addition to Exclusive Lasso. In some iterations, Group Lasso selects both clinical and gene expression variables, but this occurs less than 25\% of the time (see Figure~\ref{Ravi:Fig4}). More frequently, Group Lasso selects only variables from the gene expression group. Group Lasso also has many unique variables that are not selected by any other models, as it selects all variables from a group when that group is chosen. IPF-Lasso selects variables that are shared by all other models, with no unique variables appearing in the top 10 most frequently selected. However, no biomarker is selected more than 10 times out of 100 subsamples. Lasso also fails to select any unique or clinical features, with 5 of the top 10 features being shared by other models.

\section{Conclusion}\label{Ravi:Sec5}
Variable selection plays a critical role in high-dimensional biological datasets. Time-to-event prediction improves when redundant and non-informative features are filtered out, leading to better runtime efficiency and interpretability. However, most filter and prediction methods fail to account for the intricate grouping structure of biological data. Studies suggest that predictive performance improves when clinical variables are prioritized \citep{herrmann2021large}. However, due to their low dimensionality, clinical variables are often overshadowed by the vast number of gene expression features, particularly when using standard Lasso regularization. To ensure proper representation of low-dimensional clinical variables, we propose using Exclusive Lasso in Cox PH regression models.

The Exclusive Lasso penalty combines the \( L_1 \)-norm within groups to enforce sparsity among highly correlated features and the \( L_2 \)-norm between groups to ensure all groups are represented. This approach prevents low-dimensional groups from being overlooked while selecting the most relevant variables within each group, even when they are highly correlated. In contrast, methods like IPF-Lasso account for the grouping structure by applying an \( L_1 \)-norm within each group but do not guarantee the selection of low-dimensional groups. Additionally, a major drawback of IPF-Lasso is the need to specify a set of penalty factors or weights for each group, which, although potentially data-driven, is a time-consuming process.

In our simulation study, we compared the proposed Exclusive Lasso with other state-of-the-art methods that accounted for grouping structures, such as Elastic Net, Group Lasso, and IPF-Lasso. Exclusive Lasso outperformed these models in terms of selection accuracy and false discovery rate. Although its performance slightly deteriorated as the number of true variables increased, it still maintained a lower false discovery rate than the other models. It also performed well when variables were evenly distributed across groups. While IPF-Lasso achieved comparable performance, it either failed to select variables from certain groups or tended to select highly correlated variables within the same group. Group Lasso, on the other hand, performed poorly as it failed to select variables across all groups.

We analyzed the performance of the methods in a real-world Bladder cancer study. Although the methods had comparable integrated Brier scores, we observed that Exclusive Lasso achieved the best mean Brier score up to 60 months at every time interval. This may be because most methods tend to ignore clinical variables, whereas Exclusive Lasso selects them. The survival prediction and disease progression of bladder cancer are highly influenced by clinical predictors such as tumor stage and smoking status. Therefore, beyond gene selection, incorporating clinical variables into prediction models is crucial. We also found that variable selection in Exclusive Lasso was more consistent across repetitions, whereas other models selected different variables in different iterations.

Although Exclusive Lasso is highly effective in selecting variables from each group, its estimation is challenging due to the composite nature of the penalty. For future work, we are currently developing our proposed \emph{NM-$L_{1,2}$} algorithm \citep{ravi25} for Cox PH models. This method can robustly handle cases where certain groups contain no true variables. Traditional Exclusive Lasso cannot exclude a group even if it is non-informative due to the \( L_2 \)-norm between groups. To address this, we proposed an ad-hoc technique to include only informative variables. Specifically, we converted the composite norm into a differentiable norm and used a Newton-based algorithm for estimation. We showed that this approach outperforms the coordinate descent method. However, this method needs further investigation, particularly in scenarios with random allocation. Additionally, we plan to implement stability selection techniques, as proposed by \citet{meinshausen2010stability}, to enhance variable selection robustness, especially in cases involving random allocation. We also look forward to testing the method with more levels of grouping, such as multi-omics datasets, where, in addition to clinical variables, multiple layers of omics data—such as genomics, epigenomics, transcriptomics, proteomics, metabolomics, and microbiomics—are available.

\section*{}
\textbf{Funding}
This work has been supported by the Research Training Group “Biostatistical Methods for High-Dimensional Data in Toxicology” (RTG 2624, Project R2) funded by the Deutsche Forschungsgemeinschaft (DFG, German Research Foundation—Project Number 427806116).\\
\textbf{Competing interests}
The authors have no competing interests to declare that are relevant to the content of this article.

\bibliographystyle{apalike}
\bibliography{ref}
\end{document}